\documentclass[showpacs,prd,preprintnumbers,nofootinbib]{revtex4-1}%
\usepackage{amssymb}
\usepackage{bm}
\usepackage{amsmath}
\usepackage{graphicx}
\usepackage{amsfonts}
\usepackage{natbib}%
\providecommand{\U}[1]{\protect\rule{.1in}{.1in}}
\newif\ifinclurecommentaire
\inclurecommentairefalse

\begin{document}
\title{Near-BPS Skyrmions: Constant baryon density}
\author{Marc-Olivier Beaudoin and Luc Marleau}
\affiliation{D\'{e}partement de Physique, de G\'{e}nie Physique et d'Optique,
Universit\'{e} Laval, Qu\'{e}bec, QC, Canada G1K 7P4}
\date{\today}

\begin{abstract}
Although it provides a relatively good picture of the nucleons, the Skyrme
Model is unable to reproduce the small binding energy in nuclei. This suggests
that Skyrme-like models that nearly saturate the Bogomol'nyi bound may be more
appropriate since their mass is roughly proportional to the baryon number A.
For that purpose, we propose a near-BPS Skyrme Model. It consists of terms up
to order six in derivatives of the pion fields, including the nonlinear and
Skyrme terms which are assumed to be relatively small. For our special choice
of mass term, we obtain well-behaved analytical BPS-type solutions with
constant baryon density configurations, as opposed to the more complex
shell-like configurations found in most extensions of the Skyrme Model.
Fitting the four model parameters, we find a remarkable agreement for the
binding energy per nucleon B/A with respect to experimental data. These
results support the idea that nuclei could be near-BPS Skyrmions.

\end{abstract}

\maketitle

%

\inclurecommentairetrue
%

\inclurecommentairefalse

%

%

\section{\label{sec:Intro}Introduction}%

One of the most original and successful attempts to describe the low-energy
regime of the theory of strong interactions comes from an idea suggested by
Skyrme \cite{Skyrme:1958vn, *Skyrme:1961vq, *Skyrme:1961vr, *Skyrme:1962vh}
that baryons (and nuclei) are topological soliton solutions arising from an
effective Lagrangian of mesons. The proposal is supported by the work of
Witten \cite{Witten:1979kh} who realized that the large $N_{c}$ limit of QCD
points towards such an interpretation. More recently, an analysis of the low
energy hadron physics in holographic QCD \cite{Sakai:2004cn} have led to a
similar picture, i.e. the Skyrme Model. The model, in its original form,
succeeds in predicting the properties of the nucleon within a precision of
30\% \cite{Adkins:1983ya}. This is considered a rather good agreement for
model which involves only two parameters. Some attempts to improve the model
have given birth to a number of extensions or generalizations. Most of them
rely, to some extent, on our ignorance of the exact form of the low-energy
effective Lagrangian of QCD namely, the structure of the mass term
\cite{Marleau:1990nh,Bonenfant:2009zz,Kopeliovich:2005vg}, the contribution of
other vector mesons \cite{Sutcliffe:2008sk, Adkins:1983nw} or simply the
addition of higher-order terms in derivatives of the pion fields
\cite{Marleau:1990nh}.

Unfortunately, one of the recurring problems of Skyrme-like Lagrangians is
that they almost inevitably give nuclei binding energy that are too large by
at least an order of magnitude. Perhaps a better approach would be to
construct an effective Lagrangians with soliton solutions that nearly saturate
the Bogomol'nyi bound. If this indeed the case, then the classical static
energy of such BPS-Skyrmions (Bogomol'nyi-Prasad-Sommerfeld) grows linearly
with the baryon number $A$ (or atomic number) much like the nuclear mass.
Support for this idea comes from a recent result from Sutcliffe
\cite{Sutcliffe:2010et} who found that BPS-type Skyrmions seem to emerge for
the original Skyrme Model when a large number of vector mesons are added. The
additional degrees of freedom bring the mass of the soliton down to the
saturation of the Bogomol'nyi bound. A more direct approach to construct
BPS-Skyrmions was also proposed by Adam, Sanchez-Guillen, and Wereszczynski
(ASW) \cite{Adam:2010fg}. Their prototype model consists of only two terms:
one of order six in derivatives of the pion fields \cite{Jackson:1985yz} and a
second term, called the potential, which is chosen to be the customary mass
term for pions in the Skyrme Model \cite{Adkins:1983hy}. The model leads to
BPS-type compacton solutions with size and mass growing as $A^{\frac{1}{3}}$
and $A$ respectively, a result in general agreement with experimental
observations. However, the connection between the ASW model and pion physics,
or the Skyrme Model, is more obscure due to the absence of the nonlinear
$\sigma$ and so-called Skyrme terms which are of order 2 and 4 in derivatives, respectively.

Pursuing in this direction, some of us
\cite{Bonenfant:2010ab,Bonenfant:2012kt} reexamined a more realistic
generalization of the Skyrme Model which includes terms up to order six in
derivatives \cite{Jackson:1985yz} considering the regime where the nonlinear
$\sigma$ and Skyrme terms are are small perturbations, refered in what follows
as the near-BPS Skyrme Model . In that limit, it is possible, given an
appropriate choice of potential, to find well-behaved analytical solutions for
the static solitons in that approximation. Since they saturate the Bogomol'nyi
bound, their static energy is directly proportional to $A$ and one recovers
some of the results of Ref. \cite{Adam:2010fg}. In fact, these solutions allow
computing the mass of the nuclei including static, rotational, Coulomb and
isospin breaking energies. Adjusting the four parameters of the model to fit
the resulting binding energies per nucleon with respect to the experimental
data of the most abundant isotopes leads to an impressive agreement.

These results support the idea of a BPS-type Skyrme Model as the dominant
contribution to an effective theory for the properties of nuclear matter.
However, a few issues remain to be addressed before such a model is considered
viable. One of them concerns the shape of the energy and baryon densities. As
for most extensions of the Skyrme Model, the BPS-type models in Refs.
\cite{Adam:2010fg}, \cite{Bonenfant:2010ab} and \cite{Bonenfant:2012kt}
generate compact, shell-like or gaussian-like configurations for the energy
and baryon densities, respectively, as opposed to what experimental data
suggests, i.e. almost constant densities in the nuclei. The purpose of this
work is to show that it is possible to construct an effective Lagrangian which
leads to a uniform baryon density and still preserve the agreement with
nuclear mass data. It may be noted that near-BPS Skyrme models form a much
bigger set than previously thought as suggested from the recent discovery of
topological energy bounds \cite{Harland:2013rxa, Adam:2013tga} or different
extensions \cite{Bednarski:2013yca}.

\section{\label{sec:Skyrme}The near-BPS Skyrme Model}

We consider an extension of the original Skyrme Model that consist of the
Lagrangian density%
\begin{equation}
\mathcal{L}=\mathcal{L}_{0}+\mathcal{L}_{2}+\mathcal{L}_{4}+\mathcal{L}_{6}
\label{model0to6}%
\end{equation}
with
\begin{align}
\mathcal{L}_{0}  &  =-\mu^{2}V(U)\label{L0}\\
\mathcal{L}_{2}  &  =-\alpha\ \text{Tr}\left[  L_{\mu}L^{\mu}\right]
\label{L2}\\
\mathcal{L}_{4}  &  =\beta\ \text{Tr}\left[  f_{\mu\nu}f^{\mu\nu}\right]
\label{L4}\\
\mathcal{L}_{6}  &  =-\frac{3}{2}\frac{\lambda^{2}}{16^{2}}\text{Tr}\left[
f_{\mu\nu}f^{\nu\lambda}f_{\lambda}^{\ \ \mu}\right]  \label{L6}%
\end{align}
where $L_{\mu}=U^{\dagger}\partial_{\mu}U$ is the left-handed current and we
write for simplicity, the commutators as $f_{\mu\nu}=\left[  L_{\mu},L_{\nu
}\right]  .$ Here the pion fields are represented by the $SU(2)$ matrix
$U=\phi_{0}+i\tau_{i}\phi_{i}$ and obey the nonlinear condition $\phi_{0}%
^{2}+\phi_{i}^{2}=1$. The subscript $i$ in $\mathcal{L}_{i}$ denotes to the
number of derivatives of the pion fields which determines how each term
changes with respect to a scale transformation.

In the original Skyrme Model, only the nonlinear $\sigma$ term, $\mathcal{L}%
_{2},$ and the Skyrme term, $\mathcal{L}_{4},$ contribute. This implies that
$\alpha,\beta>0$ otherwise the static solution would not be stable against
scale transformations. A mass term --- or potential term --- $\mathcal{L}%
_{0},$ is often added to take into account chiral symmetry breaking so as to
generate a pion mass term for small fluctuations of the chiral field in
$V(U)$. We shall analyze this term in more details in the coming sections but,
as it turns out, the choice of potential $V(U)$ will have a direct bearing on
the form of the solutions and on the predictions of our model. Finally, the
term of order six in derivatives of the pion fields, $\mathcal{L}_{6},$ is
equivalent to $\mathcal{L}_{J6}=-\varepsilon_{J6}\mathcal{B}^{\mu}%
\mathcal{B}_{\mu}$ with $\varepsilon_{J6}=9\pi^{4}\lambda^{2}/4$ that was
first proposed by Jackson et al.\ \cite{Jackson:1985yz} to take into account
$\omega$-meson interactions. Here, $\mathcal{B}^{\mu}$ stands for the
topological current density
\begin{equation}
\mathcal{B}^{\mu}=\frac{\epsilon^{\mu\nu\rho\sigma}}{24\pi^{2}}\text{Tr}%
\left(  L_{\nu}L_{\rho}L_{\sigma}\right)  .
\end{equation}
The constants $\mu,$ $\alpha$, $\beta,$ and $\lambda$ are left as free
parameters although we shall focus on the regime where $\alpha$ and $\beta$
are relatively small, i.e. in the limit where the solutions remain close to
that of the BPS-solitons.

It is well known that setting the boundary condition for $U$ at infinity to a
constant in order to get finite energy solutions for the Skyrme fields also
characterizes such solutions by a conserved topological charge which Skyrme
identified as the baryon number $\mathcal{B}$ (or mass number $A$ in the
context of nuclei)
\begin{equation}
\mathcal{B}=\int d^{3}r\mathcal{B}^{0}=-\frac{\epsilon^{ijk}}{24\pi^{2}}\int
d^{3}r\text{Tr}\left(  L_{i}L_{j}L_{k}\right)  . \label{Bint}%
\end{equation}
Note that the static energy arising from $\mathcal{L}_{6}$ corresponds to the
square of the baryon density%
\[
E_{6}=\frac{9\pi^{4}\lambda^{2}}{4}\int\left(  \mathcal{B}^{0}\left(
\mathbf{r}\right)  \right)  ^{2}d^{3}r.
\]
It is often associated to the energy that would emerge if the Skyrme field is
couple to the $\omega-$meson \cite{Zahed:1986qz}%
\[
E_{\omega}=\frac{1}{2}\frac{g_{\omega}^{2}}{4\pi}\int\mathcal{B}^{0}\left(
\mathbf{r}\right)  \frac{e^{-m_{\omega}\left\vert \mathbf{r}-\mathbf{r}%
^{\prime}\right\vert }}{\left\vert \mathbf{r}-\mathbf{r}^{\prime}\right\vert
}\mathcal{B}^{0}\left(  \mathbf{r}^{\prime}\right)  d^{3}rd^{3}r^{\prime}.
\]
where instead of following the $e^{-m_{\omega}\left\vert \mathbf{r}%
-\mathbf{r}^{\prime}\right\vert }/\left\vert \mathbf{r}-\mathbf{r}^{\prime
}\right\vert \ $law, the interaction is replaced by a $\delta-$function
$\delta^{3}\left(  \mathbf{r}-\mathbf{r}^{\prime}\right)  $.

Historically, $\mathcal{L}_{0}$ and $\mathcal{L}_{6}$ were introduced to
provide a more general effective Lagrangian than the original Skyrme Model and
indeed, the Lagrangian in (\ref{model0to6}) represents the most general
$SU(2)$ model with at most two time derivatives. Since one generally relies on
the standard Hamiltonian interpretation for the quantization procedure,
higher-order time derivatives are usually avoided. On the other hand, it
should be kept in mind that an effective theory based on the $1/N_{c}$
expansion of QCD should, in principle, include terms with higher-order
derivatives of the fields.

The model (\ref{model0to6}) has been studied rather extensively in the sector
where the values of parameters $\mu,$ $\alpha$, $\beta,$ and $\lambda$ close
to that of the original Skyrme Model \cite{Jackson:1985yz, Floratos:2001ih,
*Floratos:2001bz, *Kopeliovich:2004pd, *Kopeliovich:2005hs}. Clearly these
choices were made so that $\mathcal{L}_{2}$ and $\mathcal{L}_{4}$ would
continue to have a significant contribution to the mass of the baryons and
thereby preserve the relative successes of the Skyrme Model in predicting
nucleon properties and their link to soft-pion theorems ($\alpha$ is
proportional to the pion decay constant $F_{\pi}$). Yet this sector of the
theory fails to provide an accurate description of the binding energy of heavy nuclei.

Noting that this caveat may come from the fact that the solitons of the Skyrme
Model do not saturate the Bogomol'nyi bound, ASW proposed a toy model
\cite{Adam:2010fg} (equivalent to setting $\alpha=\beta=0)$ whose solutions
are just BPS solitons.\ In principle however, the model cannot lead to stable
nuclei since BPS-soliton masses are exactly proportional to the topological
number, so $\mathcal{B}>1$ solutions have no binding energies. A more
realistic approach was proposed in Refs. \cite{Bonenfant:2010ab,
Bonenfant:2012kt} where the Lagrangian (\ref{model0to6}) is assumed to be in
the sector where $\alpha$ and $\beta$ are relatively small, treating these two
terms as perturbations. The solutions almost saturate without reaching the
Bogomol'nyi bound so that it allows for small but non-zero binding energies.
However, in spite of a very good agreement with experimental nuclear masses,
there remain a few obstacles to the acceptance of such model. For instance,
nuclear matter is believed to be uniformly distributed inside a nucleus
whereas the solutions of the aforementioned models \cite{Adam:2010fg,
Bonenfant:2010ab, Bonenfant:2012kt} display either compact, shell-like or
gaussian-like baryon and energy densities respectively. The main purpose of
this work is to demonstrate that it is possible to construct an effective
Lagrangian which leads to a uniform densities and still preserves the
agreement with nuclear mass data.

Let us consider the static solution for $U$. It can be written in the general
form
\begin{equation}
U=e^{i\mathbf{n}\cdot\mathbf{\tau}F}=\cos F+i\mathbf{n}\cdot\mathbf{\tau}\sin
F \label{Hedgehog}%
\end{equation}
where $\mathbf{\hat{n}}$ is the unit vector
\begin{equation}
\mathbf{\hat{n}}=\left(  \sin\Theta\cos\Phi,\sin\Theta\sin\Phi,\cos
\Theta\right)
\end{equation}
and $F,\Theta,$ and $\Phi$ depend in general on the spherical coordinates
$r,\theta,$ and $\phi$.

We first consider the model in (\ref{model0to6}) in the limit where $\alpha$
and $\beta$ are small. For that purpose, we introduce the axial solutions for
the $\alpha=\beta=0$ case,%
\begin{equation}
F=F(r),\qquad\Theta=\theta,\qquad\Phi=A\phi\label{axialsolution}%
\end{equation}
where $A$ is an integer that correspond to the baryon number or mass number of
a nucleus.

A word of warning is in order here. The solution (\ref{axialsolution}) is only
one of an infinite dimensional families of solutions of the BPS model and, is
not expected to be the true minimizing solution of the static energy of the
model or, for that matter, of the total energy which includes also the
(iso)rotational energy, the Coulomb energy and an isospin symmetry breaking
term. Since $\alpha$ and $\beta$ are assumed to be small, the nonlinear
$\sigma$ and Skyrme terms are not expected to a determining factor in
minimizing the total energy. In fact, the dominant effect should come from the
repulsive Coulomb energy which would have a tendency to favor a most symmetric
configuration. Which form is the true minimizer remains an open question only
to be answered by heavy numerical calculations. In the absence of such an
analysis and for the sake of simplicity, we chose to consider ansatz
(\ref{axialsolution}) which allows to easily estimate all the contributions to
the mass of the nuclei.

From hereon, we shall use whenever possible the dimensionless variable $x=ar$
where \ $a=\left(  \mu/18A\lambda\right)  ^{1/3}$ in order to factor out the
explicit dependence on the model parameters $\mu,\alpha,\beta,$ and, $\lambda$
and baryon number $A$. In fact, most of the relevant quantities can be written
in terms of three fundamental objects
\begin{align}
\left(  \mathbf{\nabla}F\right)  ^{2}  &  =\left(  a\partial_{x}F\right)
^{2}\nonumber\\
\left(  \sin F\mathbf{\nabla}\Theta\right)  ^{2}  &  =\left(  a\frac{\sin
F}{x}\right)  ^{2}\label{gradients}\\
\left(  \sin F\sin\Theta\mathbf{\nabla\Phi}\right)  ^{2}  &  =\left(
aA\frac{\sin F}{x}\right)  ^{2}\nonumber
\end{align}

The total static energy $E_{\text{s}}$ gets a contribution from each term in
(\ref{model0to6}), respectively,
\begin{align}
E_{0}  &  =4\pi\left(  \frac{\mu^{2}}{a^{3}}\right)  I_{0}^{V}\nonumber\\
E_{2}  &  =4\pi\left(  \frac{2\alpha}{a}\right)  \left(  I_{200}^{0}%
+I_{020}^{0}+I_{002}^{0}\right) \label{Estatx}\\
E_{4}  &  =4\pi\left(  16\beta a\right)  \left(  I_{220}^{0}+I_{202}%
^{0}+I_{022}^{0}\right) \nonumber\\
E_{6}  &  =4\pi\left(  \frac{9}{16}\lambda^{2}a^{3}\right)  I_{222}%
^{0}\nonumber
\end{align}
where $I_{lmn}^{k}$ are parameter-free integrals given by
\begin{align}
I_{lmn}^{k}(z)  &  =\int_{0}^{z}dx\ x^{2}\mathcal{I}_{lmn}^{k}(x)\quad
\text{with }\quad\mathcal{I}_{lmn}^{k}(x)=x^{k}\left(  \partial_{x}F\right)
^{l}\left(  \frac{\sin F}{x}\right)  ^{m}\left(  A\frac{\sin F}{x}\right)
^{n}\label{Ilmn}\\
I_{0}^{V}  &  =\int_{0}^{\infty}x^{2}dx\ V(F)=\sum_{m}C_{m}^{V}I_{0m0}^{m}
\label{IV}%
\end{align}
and write $I_{lmn}^{k}=I_{lmn}^{k}(\infty)$ for simplicity. Note that some of
these integrals are related in our case since $\mathcal{I}_{lmn}^{k}%
=A^{n}\mathcal{I}_{l,m+n,0}^{k}$. In the last equality, we assume that one can
recast $V(F)$ as a power series of $\sin F,$ i.e. $V(F)=\sum_{m}C_{m}^{V}%
\sin^{m}F$ as suggested in Ref. \cite{Marleau:1990nh}. The terms $E_{0}$ and
$E_{6}$ are proportional to the baryon number $A$ as one expects from
solutions that saturate the Bogomol'nyi bound whereas the small perturbations
$E_{2}=A^{1/3}(a_{2}+b_{2}A^{2})$ and $E_{4}=A^{-1/3}(a_{4}+b_{4}A^{2})$ have
a more complex dependence. Part of this behavior, the overall factor
$A^{\pm1/3},$ is due to the scaling. The additional factor of $A^{2}$ comes
from the axial symmetry of the solution (\ref{axialsolution}) that can be
factored out from $I_{lm2}^{k}=A^{2}I_{l,m+2,0}^{k}.$%

The topological charge also simplifies to
\begin{equation}
A=\int d^{3}x\mathcal{B}^{0}(\mathbf{x})=-\frac{2}{\pi}I_{111}^{0}%
\end{equation}
The root mean square radius of the baryon density is given by
\begin{equation}
\left\langle r^{2}\right\rangle ^{\frac{1}{2}}=\frac{1}{2\pi a}\left(
-2I_{120}^{2}\right)  ^{1/2} \label{r2baryon}%
\end{equation}
which is consistent with experimental observation for the charge distribution
of nuclei $\left\langle r^{2}\right\rangle ^{\frac{1}{2}}=r_{0}A^{\frac{1}{3}%
}$.%

The minimization of the static energy for $\alpha=\beta=0$ leads to the
differential equation for $F:$
\begin{equation}
\frac{\sin^{2}F}{288x^{2}}\partial_{x}\left(  \frac{\sin^{2}F}{x^{2}}%
\partial_{x}F\right)  -\frac{\partial V}{\partial F}=0. \label{minimisation}%
\end{equation}
Multiplying by $\partial_{x}F,$ this expression can be integrated%
\begin{equation}
\left(  \frac{\sin^{2}F}{x^{2}}\partial_{x}F\right)  ^{2}=576V
\label{equipartition}%
\end{equation}
which leads to%
\begin{equation}
\int\frac{\sin^{2}F}{8\sqrt{V}}dF=\pm\left(  x^{3}-x_{0}^{3}\right)
\label{Fz}%
\end{equation}
where $x_{0}$ is an integration constant. Finally, the expression for $F(x)$
can be found analytically provided the integral on the left-hand side is an
invertible function of $F.$ For example, assuming that the potential may be
written in the form
\begin{equation}
\sqrt{V}=\frac{u\left(  1-u^{2}\right)  }{g^{\prime}(\sqrt{1-u^{2}})}
\label{Vgu}%
\end{equation}
where $u=\cos\left(  F/2\right)  $ and, $g^{\prime}(u)=\partial g/\partial u,$
equation (\ref{Fz}) leads to
\begin{equation}
\sqrt{1-u^{2}}=\sin\left(  F/2\right)  =g^{-1}\left(  \mp\left(  x^{3}%
-x_{0}^{3}\right)  \right)  \label{Fr}%
\end{equation}
Such solutions saturate the Bogomol'nyi bound \cite{Adam:2010fg}, so their
static energy is proportional to the baryon number $A$. One would like
ultimately to reproduce the observed structure of nuclei, i.e. a roughly
constant baryon density becoming diffuse at the nuclear surface which is
characterized by a skin constant thickness parameter. Unfortunately the chiral
angle $F$ in (\ref{Fr}) cannot reproduce this last feature since $F$\ can only
be a function of the ratio $r/A^{1/3}$. So the resulting thickness parameter
is not constant and should scale like $A^{1/3}.$

It is interesting to note that (\ref{equipartition}) implies that for the
minimum energy solutions%
\begin{equation}
V(x)=\frac{1}{576}\left(  \frac{\sin F}{x}\right)  ^{4}\left(  \partial
_{x}F\right)  ^{2}\label{Vx}%
\end{equation}
so
\[
E_{0}=4\pi\left(  \frac{\lambda\mu}{32A}\right)  I_{222}^{0}=E_{6}%
\]
where the last equality arises from Derrick scaling. Furthermore according to
(\ref{Bint}) and (\ref{Vx}), the square root of the potential
\[
\sqrt{V(x)}=-\frac{1}{24}\frac{\sin^{2}F}{x^{2}}\partial_{x}F=\frac{\pi}%
{48A}\mathcal{B}^{0}(x)
\]
where $\mathcal{B}^{0}(x)$ corresponds to the radial baryon density
$\mathcal{B}^{0}(x)=\int d\Omega\ \mathcal{B}^{0}(\mathbf{x})$. Thus, in order
to obtain a nonshell baryon density, it suffices to construct a potential $V$
that does not vanish at small $x$ or, equivalently, a solution such that
$\partial_{x}F(0)\neq0.$

Expression (\ref{Vgu}) must be used with caution: it only applies for
potentials $V$ which turn out to be function of $u$ alone or, in other words,
for potentials that depends on the real part of $\ $the pion field matrix $U$
or Tr$U.$ On the other hand, $\mathcal{L}_{0}$ in (\ref{model0to6}) needs to
be explicitly written in terms of the fields $U$. A simple but not unique
approach to construct such potential is to identify $u=\cos(F/2)$ to the
expression
\[
2U_{+}=u^{2}I
\]
where $U_{\pm}=(2I\pm U\pm U^{\dagger})/8$ and $I$ is the $2\times2$ identity
matrix. Then, a convenient expression for $V(U)$ is given by%
\begin{equation}
V(U)=\frac{16\text{Tr}\left[  U_{+}U_{-}^{2}\right]  }{\left[  g^{\prime
}\left(  \left(  \text{Tr}\left[  U_{-}\right]  \right)  ^{1/2}\right)
\right]  ^{2}}\nonumber
\end{equation}

In the context of the BPS-Skyrme Model, not only the\ potential $V$ appears as
one of the dominant term in the static energy but it is also a key ingredient
in the determination of the solution. In principle, the full effective theory
including the potential should emerge from the low-energy limit of QCD, but
apart from a few symmetry arguments, little is known on the exact form of $V$.
A most simple expression for $V$ that reads
\begin{equation}
V_{\text{ASW}}(U)=-\text{Tr}\left[  U_{-}\right]  =1-u^{2} \label{V1mcos}%
\end{equation}%

was first proposed by Adkins et al. \cite{Adkins:1983hy} and served as an
additional term to the original Skyrme Lagrangian. Its main purpose was to
recover the chiral symmetry breaking pion mass term $-\frac{1}{2}m_{\pi}%
^{2}\mathbf{\pi}\cdot\mathbf{\pi}$ \ in the limit of small pion field
fluctuations $U=\exp(2i\tau_{a}\pi_{a}/F_{\pi}).$ It is sometimes useful to
recast the potential in the form \cite{Marleau:1990nh}
\begin{equation}
\mu^{2}V=\sum_{k=1}^{4}C_{k}\text{Tr}\left[  2I-U^{k}-U^{\dagger k}\right]
\label{mumpi}%
\end{equation}
Taking the limit of small pion field fluctuations, this allows fixing the
parameter $\mu$ in terms of the pion mass $m_{\pi}$ through the relation
\[
\sum_{k=1}^{\infty}k^{2}C_{k}=-\frac{m_{\pi}^{2}F_{\pi}^{2}}{16}.
\]
The choice of potential (\ref{V1mcos}) corresponds to the choice
$g(u)=u^{3}/3$ in (\ref{Fr}) and solving for $F$ leads to the BPS-compacton
solution of ASW \cite{Adam:2010fg}:
\begin{equation}
F_{\text{ASW}}(x)=\left\{
\begin{tabular}
[c]{lll}%
$2\arccos\left(  3^{1/3}x\right)  $ & $\qquad\text{for}$ & $x\in\left[
0,3^{-1/3}\right]  $\\
$0$ & $\qquad\text{for}$ & $x\geq3^{-1/3}$%
\end{tabular}
\ \ \ \ \ \ \ \ \ \ \ \ \ \ \ \ \ \ \ \text{ }\right.
\end{equation}
Note here that $\partial_{x}F(x)$ diverges as $\ x\rightarrow3^{-1/3}$ which
implies that $E_{2}$ and $E_{4}$ are not well defined. Unfortunately, this
solution as well as those arising from other similar models \cite{Adam:2012md}
saturate the Bogomol'nyi bound and as such, they give no binding energies for
the classical solitons with $B>1$.

Several alternatives to (\ref{V1mcos}) have also been proposed
\cite{Marleau:1990nh,Kopeliovich:2005vg} but recently, the major role played
by the potential in the predictions for BPS-Skyrme Models was realized and it
has led to a few interesting cases:

\begin{itemize}
\item One such example is a potential based on Ref. \cite{Bonenfant:2010ab}
\[
V_{\text{BoM}}(U)=-8\text{Tr}\left[  U_{+}U_{-}^{3}\right]
\]%

which correspond to the choice $-C_{1}=C_{2}=C_{3}=4C_{4}=\mu^{2}/128$ and
$C_{k>4}=0$ in (\ref{mumpi}). It leads to well-behaved solutions
\begin{equation}
F_{\text{BoM}}(x)=\pi\mp2\arccos\left[  \exp\left(  -x^{3}\right)  \right]
\label{exp3}%
\end{equation}
where $\partial_{x}F$ remains negative and finite for all $x.$ In order to set
the baryon number to $A,$ the boundary conditions are chose to be $F(0)=\pm
\pi$ and $F(\infty)=0$ for positive and negative baryon number respectively.
Note that the exponential fall off of $F$ at large $x$ prevents some
quantities such as the moments of inertia from becoming infinite. However,
$\partial_{x}F(x)$ vanishes at $x=0$ and so does the baryon density, leading
to an unsatisfactory shell-like configuration.

\item In that regard, a solution similar to that proposed in Ref.
\cite{Bonenfant:2012kt} seems more appropriate
\begin{equation}
F_{\text{BHM}}(x)=\pi\mp2\arccos\left[  \exp\left(  -x^{2}\right)  \right]
\end{equation}
since it possesses the kind of non-shell like baryon density configurations
observed in nature. It emerges from the potential of the form
\[
V_{\text{BHM}}(U)=-\frac{64}{9}\frac{\text{Tr}\left[  U_{+}U_{-}^{3}\right]
}{\ln\left(  \text{Tr}\left[  U_{-}\right]  \right)  }%
\]%

\end{itemize}

\begin{figure}[ptbh]
\centering\includegraphics[width=0.65\textwidth]{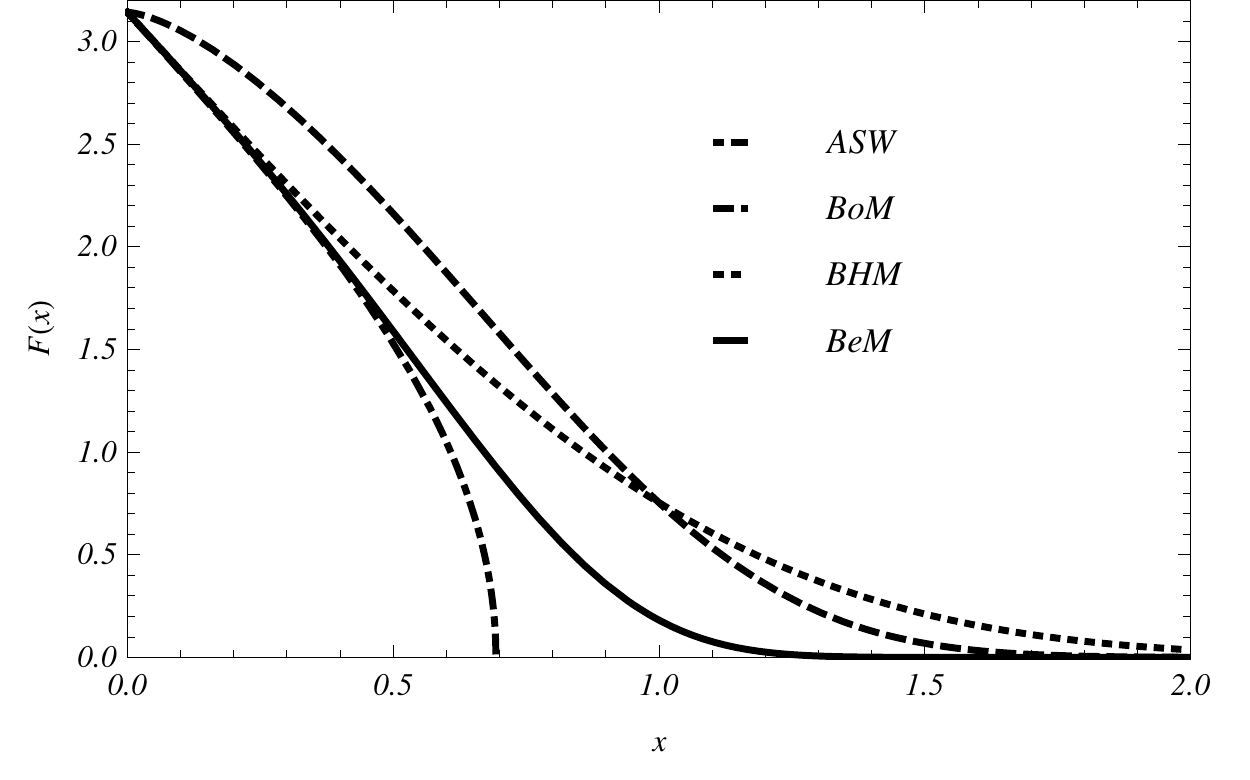}\caption{Profile
$F(x)$ for models ASW (dotdashed), BoM (dashed), BHM (dotted) and BeM
(solid).\qquad\qquad\qquad\qquad\qquad}%
\label{F}%
\end{figure}

\begin{figure}[ptbh]
\centering\includegraphics[width=0.65\textwidth]{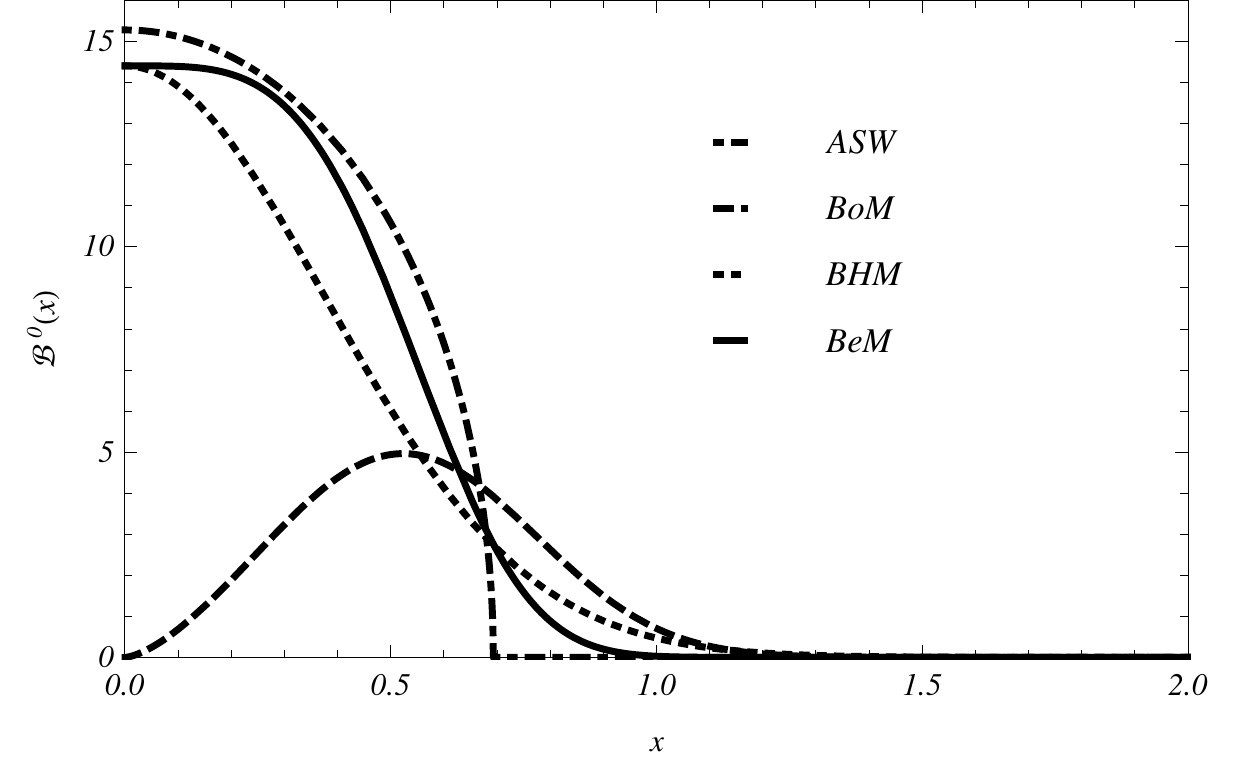}\caption{Radial
baryon density $B(x)$ for models ASW (dotdashed), BoM (dashed), BHM (dotted)
and BeM (solid).}%
\label{B}%
\end{figure}

These models display compact, shell-like or gaussian-like baryon and energy
densities (see Figs. \ref{F} and \ref{B}). However here, we shall demonstrate
that it is possible to construct an effective Lagrangian which leads to a
uniform baryon density and still preserves and even improves the agreement
with nuclear mass data.

If we assume for now that the observed baryon density can be appropriately
approximated by the parametrization $\rho_{B}(r,A)$ then, one is looking for a
solution for $F(r)$ such that%
\begin{equation}
\rho_{B}(r,A)=-\frac{A}{2\pi^{2}}\frac{\sin^{2}F}{r^{2}}F^{\prime}
\label{rhorn}%
\end{equation}
Separating variables and integrating both sides of the equation
\[
-\frac{2\pi^{2}}{A}r^{2}\rho_{B}(r,A)dr=\sin^{2}FdF
\]
we get the expression of the form
\begin{equation}
F(r)=G^{-1}(Z(r)) \label{Fremp}%
\end{equation}
where%
\begin{align*}
G(F)  &  \equiv\frac{1}{2}F-\frac{1}{4}\sin2F\\
Z(r)  &  =-\frac{2\pi^{2}}{A}\int r^{2}\rho_{B}(r,A)dr
\end{align*}
In order to be consistent, the boundary conditions for $Z$ must obey
$Z(\infty)-Z(0)=-\pi/2.$ Matching expressions (\ref{Fr}) and (\ref{Fremp})
then provides an approach to construct a model, i.e. to choose a potential
$V$, that reproduces the empirical baryon density $\rho_{B}$. Again we stress
that our model leads to BPS-Skyrmions with a profile $F$ that must be a
function of the ratio $r/A^{1/3}.$\ Unfortunately, this excludes most
parametrizations in the literature, for example, densities such as the
2-parameter Fermi or Wood-Saxon form%

\[
\rho_{B}^{\text{2pF}}(r)=\rho_{0}\frac{1+e^{-c/\tau}}{1+e^{\left(  r-c\right)
/\tau}}%
\]
since they tend to reproduce two empirical observations: (a) a baryon density
that is roughly constant for all nuclei up to their boundary where (b) it is
suppressed within a thickness $t\approx4.4\tau$ that is practically constant.
The last behavior is inconsistent with the $r/A^{1/3}$ dependence of $F.$

Let us instead construct our model by modifying the gaussian-like profile
$F_{\text{BHM}}(x)$ in such a way that baryon density $\mathcal{B}^{0}(x)$ is
approximately constant. The solution $F_{\text{BHM}}(x)$ leads to a nonshell
baryon density but it falls off too rapidly. In order to suppress this
behavior we propose a solution of the form (see Figs. \ref{F} and \ref{B})
\begin{equation}
F_{\text{BeM}}(x)=\pi\mp2\arccos\left[  \exp\left(  -x^{2}-a_{4}x^{4}\right)
\right]  \label{FBeM}%
\end{equation}
and fix the coefficient $a_{4}=7/5$ by setting to zero the first coefficient
of the series expansion of $\mathcal{B}^{0}(x)$ near $x=0$. (Note that we
could, in principle, extend this procedure by changing the argument of the
exponential to a truncated series $X(x)=x^{2}+\sum_{i=2}^{N}a_{2i}x^{2i}$.
Imposing that the density remains constant further from the core would require
to set $a_{6}=1384/525,$ $a_{8}=6302/1125,$ and so on.). It is easy to find a
potential that would allow such a solution%
\[
V_{\text{BeM}}(U)=\frac{1792}{45}\text{Tr}\left[  U_{+}U_{-}^{3}\right]
\frac{\left(  1-\left(  14/5\right)  \ln\left(  \text{Tr}\left[  U_{-}\right]
\right)  \right)  }{1-\sqrt{1-\left(  14/5\right)  \ln\left(  \text{Tr}\left[
U_{-}\right]  \right)  }}%
\]
Note that in the limit of small pion field fluctuations $U=\exp(2i\tau_{a}%
\pi_{a}/F_{\pi})$, the potential has no quadratic term in the pion field i.e.
the pion mass remains zero in this model.\

where the last result is obtained assuming the axial solution
(\ref{axialsolution}).%

Using the profile $F$ in (\ref{FBeM}), the static energy in (\ref{Estatx}) can
be calculated. Recalling that $I_{lmn}^{k}=A^{n}I_{l,m+n,0}^{k}$ for the form
of axial solution at hand, we need to evaluate numerically only four
parameter-free integrals:
\begin{align*}
I_{200}^{0}  &  =2.68798\qquad I_{020}^{0}=0.48504\qquad I_{220}^{0}=5.13755\\
I_{040}^{0}  &  =1.88156\qquad I_{240}^{0}=20.27798.
\end{align*}

In order to represent physical nuclei, we have taken into account their
rotational and isorotational degrees of freedom and quantize the solitons. The
standard procedure is to use the semiclassical quantization which is described
in the next section.

\section{\label{sec:Quantization}Quantization}

Skyrmions are not pointlike particles so we resort to a semiclassical
quantization method which consists in adding an explicit time dependence to
the zero modes of the Skyrmions and applying a time-dependent (iso)rotations
on the Skyrme fields by $SU(2)$ matrix $A_{1}(t)$ and $A_{2}(t)$
\begin{equation}
\tilde{U}(\mathbf{r},t)=A_{1}(t)U(R(A_{2}(t))\mathbf{r})A_{1}^{\dag}(t)
\end{equation}
where $R(A_{2}(t))$ is the associated $SO(3)$ rotation matrix. The approach
assumes that the Skyrmion behaves as a rigid rotator. Upon insertion of this
ansatz in the time-dependent part of the full Lagrangian (\ref{model0to6}), we
can write the (iso)rotational Lagrangian as
\begin{equation}
\mathcal{L}_{\text{r}}=\frac{1}{2}a_{i}U_{ij}a_{j}-a_{i}W_{ij}b_{j}+\frac
{1}{2}b_{i}V_{ij}b_{j},
\end{equation}
where $a_{k}=-i$Tr$\tau_{k}A_{1}^{\dag}\dot{A}_{1}$ and $b_{k}=i$Tr$\tau
_{k}\dot{A}_{2}A_{2}^{\dag}$

The moment of inertia tensors $U_{ij}$ are given by%
\begin{align}
U_{ij}  &  =\int d^{3}r\ \mathcal{U}_{ij}=-\frac{1}{a}\int d^{3}x\left[
\frac{2\alpha}{a^{2}}\text{Tr}\left(  T_{i}T_{j}\right)  \right. \nonumber\\
&  +4\beta\text{Tr}\left(  \left[  L_{p},T_{i}\right]  \left[  L_{p}%
,T_{j}\right]  \right) \nonumber\\
&  +\left.  \frac{9\lambda^{2}}{16^{2}}a^{2}\text{Tr}\left(  \left[
T_{i},L_{p}\right]  \left[  L_{p},L_{q}\right]  \left[  L_{q},T_{j}\right]
\right)  \right]  \label{MInertia}%
\end{align}
where $T_{i}=iU^{\dagger}\left[  \frac{\tau_{i}}{2},U\right]  $. The
expressions for $W_{ij}$ and $V_{ij}$ are similar except that the
isorotational operator $T_{i}$ is replaced by a rotational analog
$S_{i}=-\epsilon_{ikl}x_{k}L_{l}$ as follows:
\begin{align}
W_{ij}  &  =\int d^{3}r\ \mathcal{W}_{ij}=\int d^{3}r\ \mathcal{U}_{ij}%
(T_{j}\rightarrow S_{j})\label{Wij}\\
V_{ij}  &  =\int d^{3}r\ \mathcal{V}_{ij}=\int d^{3}r\ \mathcal{U}_{ij}%
(T_{j}\rightarrow S_{j},T_{i}\rightarrow S_{i}). \label{Vij}%
\end{align}
Following the calculations in \cite{Bonenfant:2010ab} for axial solution of
the form (\ref{axialsolution}), we find that all off-diagonal elements of the
inertia tensors vanish.

Furthermore, one can show that $U_{11}=U_{22}$ and $U_{33}$ can be obtained by
setting $A=1$ in the expression for $U_{11}$. Similar identities hold for
$V_{ij}$ and $W_{ij}$ tensors. Finally the general expressions for the moments
of inertia coming from each pieces of the Lagrangian read%

\begin{align}
U_{11}  &  =\frac{4\pi}{3a}\left(  \frac{8\alpha}{a^{2}}I_{020}^{2}%
+16\beta\left(  4I_{220}^{2}+3I_{022}^{2}+I_{040}^{2}\right)  +\frac
{9\lambda^{2}a^{2}}{16}\left(  3I_{222}^{2}+I_{240}^{2}\right)  \right)
\label{U11}\\
V_{11}  &  =\frac{4\pi}{3a}\left(  \frac{2\alpha}{a^{2}}\left(  I_{002}%
^{2}+3I_{020}^{2}\right)  +16\beta\left[  \left(  I_{202}^{2}+3I_{220}%
^{2}\right)  +4I_{022}^{2}\right]  +\frac{9\lambda^{2}a^{2}}{4}I_{222}%
^{2}\right)  \label{V11}%
\end{align}
where due to the axial form of our solution, we can extract an explicit
dependence on $A$ through the relation $I_{lmn}^{k}=A^{n}I_{l,m+n,0}^{k}.$

The axial symmetry of the solution imposes the constraint $L_{3}+AK_{3}=0$
which is simply the statement that a spatial rotation by an angle $\theta$
about the axis of symmetry can be compensated by an isorotation of $-A\theta$
about the $\tau_{3}$ axis. It follows from expressions (\ref{MInertia}%
)-(\ref{Vij}) that $W_{11}=W_{22}=0$ for $\left\vert A\right\vert \geq2$ and
$A^{2}U_{33}=AW_{33}=V_{33}$. Otherwise, for $\left\vert A\right\vert =1$, the
solution have spherical symmetry and
\begin{equation}
W_{11}=\frac{4\pi}{3a}\left(  \frac{8\alpha}{a^{2}}I_{020}^{2}+64\beta\left(
I_{220}^{2}+I_{040}^{2}\right)  +\frac{9\lambda^{2}a^{2}}{4}I_{240}%
^{2}\right)  . \label{W11}%
\end{equation}
where here $A=1$ in $a$ as well.%

The general form of the rotational Hamiltonian is given by
\cite{Houghton:2005iu}
\begin{equation}
H_{\text{r}}=H_{\text{r}}=\frac{1}{2}%
{\displaystyle\sum\limits_{i=1,2,3}}
\left[  \frac{\left(  L_{i}+W_{ii}\frac{K_{i}}{U_{ii}}\right)  ^{2}}%
{V_{ii}-\frac{W_{ii}^{2}}{U_{ii}}}+\frac{K_{i}^{2}}{U_{ii}}\right]
\label{Hrot}%
\end{equation}
where ($K_{i}$) $L_{i}$ the body-fixed (iso)rotation momentum canonically
conjugate to $(a_{i}$) $b_{i}$. It is also easy to calculate the rotational
energies for nuclei with winding number $\left\vert A\right\vert \geq2$%
\begin{equation}
H_{\text{r}}=\frac{1}{2}\left[  \frac{\mathbf{L}^{2}}{V_{11}}+\frac
{\mathbf{K}^{2}}{U_{11}}+\xi K_{3}^{2}\right]
\end{equation}
with%
\[
\xi=\frac{1}{U_{33}}-\frac{1}{U_{11}}-\frac{A^{2}}{V_{11}}%
\]
These momenta are related to the usual space-fixed isospin ($\mathbf{I}$) and
spin ($\mathbf{J}$) by the orthogonal transformations
\begin{align}
I_{i}  &  =-\frac{1}{2}\text{Tr}\left(  \tau_{i}A_{1}\tau_{j}A_{1}^{\dag
}\right)  K_{j}=-R(A_{1})_{ij}K_{j},\label{eq:I}\\
J_{i}  &  =-\frac{1}{2}\text{Tr}\left(  \tau_{i}A_{2}\tau_{j}A_{2}^{\dag
}\right)  ^{\text{T}}L_{j}=-R(A_{2})_{ij}^{\text{T}}L_{j}. \label{eq:J}%
\end{align}

According to (\ref{eq:I}) and (\ref{eq:J}), we see that the Casimir invariants
satisfy $\mathbf{K}^{2}=\mathbf{I}^{2}$ and $\mathbf{L}^{2}=\mathbf{J}^{2}$ so
the rotational Hamiltonian is given by
\begin{equation}
H_{\text{r}}=\frac{1}{2}\left[  \frac{\mathbf{J}^{2}}{V_{11}}+\frac
{\mathbf{I}^{2}}{U_{11}}+\xi K_{3}^{2}\right]  . \label{Erot}%
\end{equation}
We are looking for the lowest eigenvalue of $H_{\text{r}}$ which depends on
the dimension of the spin and isospin representation of the nucleus eigenstate
$|N\rangle\equiv|i,i_{3},k_{3}\rangle|j,j_{3},l_{3}\rangle$. For $\alpha
=\beta=0,$ we can show that $\xi$ is negative and we shall assume that this
remains true for small values of $\alpha$ and $\beta$. Then, for a given spin
$j$ and isospin $i$, $\kappa$ must take the largest possible eigenvalue
$k_{3}.$ Note that $\mathbf{K}^{2}=\mathbf{I}^{2}$ and $\mathbf{L}%
^{2}=\mathbf{J}^{2},$ so the state with highest weight is characterized by
$k_{3}=i$ and $l_{3}=j.$ Furthermore, since nuclei are build out of $A$
fermions, the eigenvalues $k_{3}$ are limited to $k_{3}\leq i\leq A/2.$ On the
other hand, the axial symmetry of the static solution (\ref{axialsolution})
implies that $k_{3}=-l_{3}/A\leq j/A$ where $j\leq A/2$ as well$.$ In order to
minimize $H_{\text{r}}$, we need the largest possible eigenvalue $k_{3}$, so
for even $A$ nuclei, $\kappa$ must be an integer such that
\[
\kappa=\max(\left\vert k_{3}\right\vert )=\min\left(  i,\left[  j/A\right]
\right)  .
\]
Similarly for odd nuclei, $\left\vert k_{3}\right\vert $ must be a positive
half-integer so the only possible value is
\[
\kappa=\min\left(  i,\left[  j/A\right]  +\frac{1}{2}\right)  =\frac{1}{2}%
\]
This last relation only holds for the largest possible spin eigenstate $j=A/2$
which is not the most stable in general and so it signals that the ansatz
(\ref{axialsolution}) may not be the most appropriate for odd nuclei. The
axial symmetry may however be only marginally broken if we consider the odd
nucleus as a combination of an additional nucleon with an even nucleus
especially for large nuclei. Nonetheless, we shall retain the ansatz
(\ref{axialsolution}) for both even and odd nuclei and choose the largest
possible eigenvalue $k_{3}$ for the most stable isotopes as%
\begin{equation}
\kappa=\left\{
\begin{tabular}
[c]{l}%
$0\qquad$for $A=$ even\\
$\frac{1}{2}\qquad$for $A=$ odd
\end{tabular}
\ \ \ \ \ \ \ \ \ \ \ \ \ \ \ \ \ \ \ \ \right.  . \label{kappa}%
\end{equation}

The lowest eigenvalue of the rotational Hamiltonian $H_{\text{r}}$ for a
nucleus is then given by \cite{Bonenfant:2010ab}
\begin{equation}
E_{\text{r}}=\frac{1}{2}\left[  \frac{j(j+1)}{V_{11}}+\frac{i(i+1)}{U_{11}%
}+\xi\kappa^{2}\right]  \label{Erotijk}%
\end{equation}
The spins of the most abundant isotopes are well known. This is not the case
for the isospins so we resort to the usual assumption that the most abundant
isotopes correspond to states with lowest isorotational energy. Since
$i\geq\left\vert i_{3}\right\vert $, the lowest value that $i$ can take is
simply $\left\vert i_{3}\right\vert $ where $i_{3}=Z-A/2.$ For example, the
nucleon and deuteron rotational energy reduces respectively to%
\begin{align}
E_{\text{r}}^{N}  &  =\frac{3}{8U_{11}}\quad A=1,\ j=i=\kappa=1/2\text{ }\\
E_{\text{r}}^{D}  &  =\frac{1}{V_{11}}\quad A=2,\ j=1,\ i=\kappa=0\text{ }%
\end{align}

The explicit calculations of the rotational energy of each nucleus then
require the numerical evaluation of the following four parameter-free
integrals in (\ref{U11}), (\ref{V11}) and (\ref{W11}) which, in our model,
turn out to be
\begin{align*}
I_{020}^{2}  &  =0.142868\qquad I_{220}^{2}=1.43364\\
I_{040}^{2}  &  =0.352712\qquad I_{240}^{2}=3.94598.
\end{align*}

So far, both contributions to the mass of the nucleus, $E_{\text{s}}$ and
$E_{\text{r}},$ are charge invariant. Since this is a symmetry of the strong
interaction, it is reflected in the construction of the Lagrangian
(\ref{model0to6}) and one expects that the two terms form the dominant portion
of the mass. However, isotope masses differ by a few percent so this symmetry
is broken for physical nuclei. In the next section, we consider two additional
contributions to the mass, the Coulomb energy associated with the charge
distribution inside the Skyrmion and an isospin breaking term\ that may be
attributed to the up and down quark mass difference.

\section{\label{sec:Coulomb}Coulomb energy and isospin breaking}

The electromagnetic and isospin breaking contributions to the mass have been
thoroughly studied for $A=1$, mostly in the context of the computation of the
proton-neutron mass difference \cite{Durgut:1985mu, *Kaelbermann:1986ne,
*Ebrahim:1987mu,*Jain:1989kn, *Weigel:1989eb,Rathske:1988qt,Meissner:2009hm},
but are usually neglected, to a first approximation, for higher $A$ since they
are not expected to overcome the large binding energies predicted by the
model. There are also practical reasons why they are seldom taken into
account. The higher baryon number configurations of the original Skyrme Model
are nontrivial (toroidal shape for $A=2$, tetrahedral for $A=3$, etc.) and
finding them exactly either requires heavy numerical calculations (see for
example \cite{Longpre:2005fr}) or some kind of clever approximation like
rational maps \cite{Houghton:1997kg}. In our case however, we are interested
in a precise calculation of the nuclei masses and an estimate of the Coulomb
energy is desirable, and even more so in our model which generates nonshell
configurations. It turns out that the axial symmetry of the solution and the
relatively simple form of the chiral angle $F(r)$ in (\ref{FBeM}) simplify the
computation of the Coulomb energy.

Let us first consider the charge density inside Skyrmions. Following Adkins et
al. \cite{Adkins:1983ya}, we write the electromagnetic current
\begin{equation}
J_{EM}^{\mu}=\frac{1}{2}\mathcal{B}^{\mu}+J_{V}^{\mu3},
\end{equation}
with $\mathcal{B}^{\mu}$ the baryon density and $J_{V}^{\mu3}$ the vector
current density. The conserved electric charge is given by
\begin{equation}
Z=\int d^{3}rJ_{EM}^{0}=\int d^{3}r\left(  \frac{1}{2}\mathcal{B}^{0}%
+J_{V}^{03}\right)  \label{charge}%
\end{equation}
The vector current is then defined as the sum of the left and right handed
currents
\[
J_{V}^{\mu i}=J_{R}^{\mu i}+J_{L}^{\mu i}%
\]
which are invariant under $SU(2)_{L}\otimes SU(2)_{R}$ transformations of the
form $U\rightarrow LUR^{\dagger}.$ More explicitly, we get%
\begin{equation}
J_{V}^{0i}=-\frac{1}{2}\{R(A_{1})_{ij},\left(  \mathcal{U}_{jk}a_{k}%
-\mathcal{W}_{jk}b_{k}\right)  \}\label{J3V}%
\end{equation}
where $\mathcal{U}_{ij}$ and $\mathcal{W}_{ij}$ are the moment of inertia
densities in (\ref{MInertia})-(\ref{Vij}). The calculations of the Coulomb
energy here follows that in \cite{Adam:2013tda, *Adam:2013wya}; it differs
that from Ref. \cite{Bonenfant:2012kt} where only the body-fixed charge
density was considered. The anticommutator is introduced to ensure that
$J_{V}^{0i}$ is a Hermitian operator. In the quantized version, $a_{j}\ $and
$b_{j}$ are expressed in terms of the conjugate operators $K_{i}$ and $L_{i}.$
Here we only need the relation
\[
K_{i}=U_{ij}a_{j}-W_{ij}b_{j}%
\]
The solution is axially symmetric then the off-diagonal elements of $U_{ij}$
and $W_{ij}$ vanish, $W_{11}=W_{22}=0$ for $\left\vert A\right\vert \geq2$ and
$AU_{33}=W_{33}$. Then have
\[
a_{1}=\frac{K_{1}}{U_{11}},\qquad a_{2}=\frac{K_{2}}{U_{22}},\qquad
a_{3}=\frac{K_{3}}{U_{33}}+Ab_{3}%
\]
Inserting $a_{i}$ in (\ref{J3V}), the isovector electric current density
reduces to
\[
J_{V}^{03}=-\frac{1}{2}\{R(A_{1})_{3i},\frac{\mathcal{U}_{ii}}{U_{ii}}K_{i}\}
\]
where $\mathcal{U}_{ii}/U_{ii}$ may be interpreted here as a normalized moment
of inertia density for the $i^{\text{th}}$ component of isospin in the
body-fixed frame. The expectation value $R(A_{1})_{31}K_{1}\ $and
$R(A_{1})_{32}K_{2}$ for eigenstate $|N\rangle=|i,i_{3},k_{3}\rangle
|j,j_{3},l_{3}\rangle$ are equal so that we may simplify%

\begin{equation}
\left\langle N\right\vert J_{V}^{03}|N\rangle=\frac{\mathcal{U}_{11}%
+\mathcal{U}_{22}}{2U_{11}}i_{3}+\left[  \frac{\mathcal{U}_{11}+\mathcal{U}%
_{22}}{2U_{11}}-\frac{\mathcal{U}_{33}}{U_{33}}\right]  \left\langle
N\right\vert R(A_{1})_{33}K_{3}|N\rangle\label{J03VN}%
\end{equation}
where we have used relation (\ref{eq:I}). The moment of inertia density are
given by
\begin{align}
\mathcal{U}_{11}+\mathcal{U}_{22} &  =4\alpha\mathcal{I}_{020}^{2}(1+\cos
^{2}\theta)+32\beta a^{2}\left(  \mathcal{I}_{220}^{2}(1+\cos^{2}%
\theta)+\mathcal{I}_{040}^{2}\left(  A^{2}+\cos^{2}\theta\right)  \right)
\nonumber\\
&  +\frac{9\lambda^{2}}{8}a^{4}\mathcal{I}_{240}^{2}\left(  A^{2}+\cos
^{2}\theta\right)  \label{U11dens}\\
\mathcal{U}_{33} &  =\left(  4\alpha\mathcal{I}_{020}^{2}+32\beta a^{2}\left(
\mathcal{I}_{220}^{2}+\mathcal{I}_{040}^{2}\right)  +\frac{9\lambda^{2}}%
{8}a^{4}\mathcal{I}_{240}^{2}\right)  \sin^{2}\theta\label{U33dens}%
\end{align}
The expression in brakets in equation (\ref{J03VN}) integrates to zero so that
one recovers the relation $Z=A/2+i_{3}$ as expected. But while it does not
contribute to the total charge, the charge density is not zero everywhere. Let
us examine this contribution in more details. Since the electric charge does
not depend on the angular momentum, we can limit our analysis to the isospin
wavefunctions. Following Adkins \cite{Adkins:1987kj} we write the
wavefunctions $\left\langle A_{1}\right.  |i,i_{3},k_{3}\rangle$ in terms of
the Wigner's functions $D_{mm^{\prime}}^{n}:$
\[
\left\langle A_{1}\right.  |i,i_{3},k_{3}\rangle=\left(  \frac{2i+1}{2\pi^{2}%
}\right)  ^{1/2}D_{k_{3}i_{3}}^{i}\left(  A_{1}\right)
\]
Similarly the matrix $R(A_{1})_{33}$ corresponds to a spin zero and isospin
zero transition that can be written
\[
R(A_{1})_{33}=D_{00}^{1}\left(  A_{1}\right)
\]
The appropriate expectation value is then given by%
\begin{align*}
\left\langle i,i_{3},k_{3}\right\vert R(A_{1})_{33}K_{3}|i,i_{3},k_{3}\rangle
&  =k_{3}\int dA_{1}\left(  \frac{2i+1}{2\pi^{2}}\right)  \left(
D_{k_{3}i_{3}}^{i}\left(  A_{1}\right)  \right)  ^{\ast}D_{00}^{1}\left(
A_{1}\right)  D_{k_{3}i_{3}}^{i}\left(  A_{1}\right)  \\
&  =k_{3}\left(  -1\right)  ^{2(k_{3}+1-i)}\left\langle 1,0;i,k_{3}%
|i,k_{3}\right\rangle \left\langle 1,0;i,i_{3}|i,i_{3}\right\rangle \\
&  =\left\{
\begin{array}
[c]{c}%
\frac{i_{3}k_{3}^{2}}{i(i+1)}\qquad\text{ for }i\neq0\\
0\qquad\text{ for }i=0
\end{array}
\right.
\end{align*}
where the last two expressions on the second line are Clebsch-Gordan
coefficients. Recalling that we have imposed the condition $\left\vert
k_{3}\right\vert =\kappa=0$ or $1/2$ for even and odd nuclei respectively and
fixed the value of the isospin to $i=\left\vert i_{3}\right\vert $, we find
\begin{equation}
\rho\equiv\frac{1}{2}\mathcal{B}^{0}+\frac{\mathcal{U}_{11}+\mathcal{U}_{22}%
}{2U_{11}}i_{3}+\left[  \frac{\mathcal{U}_{11}+\mathcal{U}_{22}}{2U_{11}%
}-\frac{\mathcal{U}_{33}}{U_{33}}\right]  \frac{i_{3}\kappa^{2}}%
{i(i+1)}\label{rhocharge}%
\end{equation}
The last term drops for even nuclei ($\kappa=0$). For odd nuclei, the
cancellation in the brackets leads a relatively small contribution which is
further suppressed by the factor $\kappa^{2}/\left(  i+1\right)  $ for large
nuclei. It is indicative of the asymmetry in the moments of inertia.%

The Coulomb energy associated to a given charge distribution $\rho
(\mathbf{r})$ takes the usual form
\begin{equation}
E_{\text{C}}=\frac{1}{2}\frac{1}{4\pi}\int\rho\left(  \mathbf{r}\right)
\frac{1}{\left\vert \mathbf{r}-\mathbf{r}^{\prime}\right\vert }\rho\left(
\mathbf{r}^{\prime}\right)  d^{3}rd^{3}r^{\prime} \label{ECoulomb}%
\end{equation}
Since we have at hand an axially symmetric distribution, it is convenient to
expand $\rho(\mathbf{r})$ in terms of normalized spherical harmonics to
perform the angular integrations
\begin{equation}
\rho(\mathbf{r})=a^{3}\rho(\mathbf{x})=a^{3}\sum_{l,m}\rho_{lm}(x)Y_{l}%
^{m\ast}(\theta,\phi). \label{Ylm}%
\end{equation}
Following the approach described in \cite{Carlson:1963mr}, we define the
quantities
\begin{equation}
Q_{lm}(r)=\int_{0}^{r}d\tilde{r}\ \tilde{r}^{l+2}\rho_{lm}(\tilde{r}%
)=a^{-l}Q_{lm}(x)
\end{equation}
which, at large distance, are equivalent to a multipole moments of the
distribution. The total Coulomb energy is given by%
\[
E_{\text{C}}=\sum_{l=0}^{\infty}\sum_{m=-l}^{l}U_{lm}%
\]
where
\[
U_{lm}=\left(  2\pi\alpha_{em}\right)  a\int_{0}^{\infty}dx\ x^{-2l-2}%
|Q_{lm}(x)|^{2}\
\]

The isocalar part to the charge distribution is a spherically symmetric
contribution%
\[
\mathcal{B}^{0}(\mathbf{r})=a^{3}\mathcal{B}^{0}(\mathbf{x})=-\frac{a^{3}%
}{2\pi^{2}}\mathcal{I}_{111}^{0}(x)
\]
where $\mathcal{I}_{lmn}^{k}$ is defined in (\ref{Ilmn}). On the other hand,
the isovector contribution in (\ref{U1122dens}) possesses a simple angular
dependence so that%

the summation (\ref{Ylm}) consists of only two terms in $Y_{0}^{0\ast}$ and
$Y_{2}^{0\ast}$.

The moments $Q_{00}$ and $Q_{20}$ are then given by%
\begin{align*}
Q_{00}(x)  & =\frac{2\sqrt{\pi}}{3}\left(  -\frac{3A}{4\pi^{2}}I_{120}%
^{0}(x)+\frac{i_{3}}{a}\left(  \frac{8\alpha}{a^{2}}I_{020}^{2}(x)C_{-}%
+16\beta\left(  4I_{220}^{2}(x)C_{-}+C_{A}I_{040}^{2}(x)\right)  \right.
\right.  \\
& +\left.  \left.  \frac{9\lambda^{2}a^{2}}{16}C_{A}I_{240}^{2}(x)\right)
\right)
\end{align*}%
\[
Q_{20}(x)=\frac{4}{3}\sqrt{\frac{\pi}{5}}\frac{i_{3}}{a}C_{+}\left(
\frac{2\alpha}{a^{2}}I_{020}^{4}(x)+16\beta\left(  I_{220}^{4}(x)+I_{040}%
^{4}(x)\right)  +\frac{9\lambda^{2}}{16}aI_{240}^{4}(x)\right)
\]
where
\begin{align*}
C_{\pm} &  =\frac{1+C}{U_{11}}+\frac{C}{2U_{33}}\pm\frac{3C}{2U_{33}}\\
C_{A} &  =(3A^{2}+1)\left(  \frac{1+C}{U_{11}}\right)  -\frac{4C}{U_{33}}%
\end{align*}
and $C=k_{3}^{2}/i(i+1)$. Finally, the Coulomb energy then takes the form%
\begin{equation}
E_{\text{C}}=\left(  2\pi\alpha_{em}\right)  a\int_{0}^{\infty}(Q_{00}%
^{2}x^{-4}+Q_{20}^{2}x^{-8})\ x^{2}dx\label{EC}%
\end{equation}

It is again convenient to regroup the model parameters in the dimensionless
quantity
\[
\mathbf{p}_{0}=\left[  A,C_{-}\frac{\alpha}{a^{3}}i_{3},C_{A}\frac{\beta}%
{a}i_{3},C_{-}\frac{\beta}{a}i_{3},C_{A}\lambda^{2}ai_{3}\right]
\]

\[
\mathbf{p}_{2}=C_{+}i_{3}\left[  \frac{\alpha}{a^{3}},\frac{\beta}{a}%
,\lambda^{2}a\right]
\]
such that we may write%
\begin{equation}
E_{\text{C}}=2\pi\alpha_{em}a\ \left(  p_{0}^{i}M_{00}^{ij}p_{0}^{j}+p_{2}%
^{i}M_{00}^{ij}p_{2}^{j}\right)  .
\end{equation}
Here, each element of $M_{00}^{ij}$ ($M_{20}^{ij}$) comes from squaring
$Q_{00}$ ($Q_{20}$) in (\ref{EC}) and depend only on the form of the profile
$F(x)$ and baryon number $A$ according to
\[
M_{l0}^{ij}=\int_{0}^{\infty}v_{l}^{i}v_{l}^{j}x^{-2-2l}dx
\]
where
\begin{align*}
\mathbf{v}_{0}  &  =\frac{2\sqrt{\pi}}{3}\left(  -\frac{3}{4\pi^{2}}%
I_{120}^{0}(x),8I_{020}^{2}(x),16I_{040}^{2}(x),64I_{220}^{2}(x),\frac{9}%
{16}I_{240}^{2}(x)\right) \\
\mathbf{v}_{2}  &  =\frac{4}{3}\sqrt{\frac{\pi}{5}}\left(  2I_{020}%
^{4}(x),16\left(  I_{220}^{4}(x)+I_{040}^{4}(x)\right)  ,\frac{9}{16}%
I_{240}^{4}(x)\right)
\end{align*}
For the solutions at hand (\ref{FBeM}), we get%
\[
\mathbf{M}_{00}=\left(
\begin{array}
[c]{ccccc}%
0.035244 & 0.295938 & 1.67062 & 24.5793 & 0.65734\\
0.295938 & 2.6131624 & 14.1112 & 215.6395 & 5.56078\\
1.67062 & 14.1112 & 79.5851 & 1173.4095 & 31.3461\\
24.5793 & 215.6395 & 1173.4095 & 17835.4373 & 462.538\\
0.65734 & 5.56078 & 31.3461 & 462.538 & 12.3494
\end{array}
\right)
\]%
\[
\mathbf{M}_{20}=\left(
\begin{array}
[c]{ccc}%
0.0156167 & 1.62666 & 0.126600028\\
1.62666 & 173.309 & 13.9867\\
0.126600028 & 13.9867 & 1.20944
\end{array}
\right)
\]
$\allowbreak$%

The Coulomb energy can explain part of the isotope mass differences, but it is
certainly not sufficient. For example for the nucleon, the Coulomb energy
would suggest that the neutron mass is smaller than that of the proton. Of
course, one can invoke the fact that isospin is not an exact symmetry to
improve the predictions. Several attempts have been proposed to parametrize
the isospin symmetry breaking term within the Skyrme Model
\cite{Rathske:1988qt,Meissner:2009hm}. Here we shall assume for simplicity
that this results in a contribution proportional to the third component of
isospin%
\begin{equation}
E_{\text{I}}=a_{I}i_{3} \label{EI}%
\end{equation}
where the parameter $a_{I}$ is fixed by setting the neutron-proton mass
difference to its experimental value $\Delta M_{n-p}^{\text{expt}}=1.293$ MeV.
Since both of them have the same static and rotational energies, we find%
\begin{equation}
a_{I}=\left(  E_{\text{C}}^{n}-E_{\text{C}}^{p}\right)  -\Delta M_{n-p}%
^{\text{expt}} \label{aI}%
\end{equation}
where $E_{\text{C}}^{n}$ and $E_{\text{C}}^{p}$ are the neutron and proton
Coulomb energy, respectively.

Summarizing, the mass of a nucleus reads%
\begin{equation}
E(A,i,j,k_{3},i_{3})=E_{\text{s}}(A)+E_{\text{r}}(A,i,j,k_{3})+E_{\text{C}%
}(A,i_{3})+E_{\text{I}}(A,i_{3}) \label{Etot}%
\end{equation}
where $E_{\text{s}}\,\ $is the total static energy. The prediction depends on
the parameters of the model $\mu,$ $\alpha,\beta,$ and $\lambda$ \ and the
relevant quantum numbers of each nucleus as shown in (\ref{Etot}).

\section{\label{sec:Model}Results and discussion}

The values of the parameters $\mu,$ $\alpha,\beta$ and $\lambda$\ remain to be
fixed. Let us first consider the case where $\alpha=\beta=0$. This should
provide us with a good estimate for the values of $\mu,\alpha,\beta,$ and
$\lambda$ required in the 4-parameter model (\ref{model0to6}) \ and, after
all, it corresponds to the limit where the minimization of the static energy
leads to the exact analytical BPS solution in (\ref{FBeM}). For simplicity, we
choose the mass of the nucleon and that of a nucleus$\ X$ with no
(iso)rotational energy (i.e. a nucleus with zero spin and isospin)\ as input
parameters. Neglecting for now the Coulomb and isospin breaking energies, the
mass of these two states is according to expression (\ref{Etot})
\begin{align*}
E_{N}  &  =15.92628\lambda\mu+0.026426\mu^{-1/3}\lambda^{-5/3}\\
E_{X}  &  =15.92628A\lambda\mu
\end{align*}

For example, if the nucleus $X$ is Calcium-40, a doubly magic number nucleus,
with mass $E_{\text{Ca}}=37214.7$ MeV, then solving for $\lambda$ and $\mu,$
we get the numerical values $\mu=12322.3$ MeV$^{2}$ , $\alpha=\beta=0$ and,
$\lambda=0.00474078$ MeV$^{-1}$ which we shall refer as Set~I. The masses of
the nuclei are then computed using Eq. (\ref{Etot}) which results in
predictions that are accurate to at least $0.6\%$, even for heavier nuclei.
This precision is somewhat expected since the static energy of a BPS-type
solution is proportional to $A$ so if it dominates, the nuclear masses should
follow approximately the same pattern. However, the predictions remain
surprisingly good compared to that of the original Skyrme model, another
2-parameter model$.$

Perhaps even more relevant are the predictions of the binding energy per
nucleon $B/A=\left(  E-Zm_{p}-(A-Z)m_{n}\right)  /A$, in which case, the
calculation simplifies. For example, subtracting from the static energy of a
nucleus from that of its constituents we find that the binding energy does not
depend on the static energies $E_{0}$ or $E_{6},$
\begin{align*}
\Delta E_{\text{s}}  &  =AE_{\text{s}}(1)-E_{\text{s}}(A)\\
&  =4\pi\left(  A-1\right)  \left(  \frac{2\alpha}{a}\left(  I_{200}%
^{0}-\left(  A-1\right)  I_{020}^{0}\right)  -16a\beta\left(  \left(
A-1\right)  I_{220}^{0}+AI_{040}^{0}\right)  \right)
\end{align*}
whereas the contribution from $E_{\text{I}}$ simply cancels out. The dominant
contributions come from the (iso)rotational and Coulomb energy differences, respectively,%

\[
\Delta E_{\text{r}}=AE_{\text{r}}^{N}-E_{\text{r}}(A,i,j,k_{3})
\]
dominated by $AE_{\text{r}}^{N}$ for large nuclei and%
\[
\Delta E_{\text{C}}=ZE_{\text{C}}^{p}+(A-Z)E_{\text{C}}^{n}-E_{\text{C}%
}(A,i_{3})
\]
which is, of course, negative due to the repulsive nature of the Coulomb force
between nucleons.

\begin{figure}[ptbh]
\centering\includegraphics[width=0.65\textwidth]{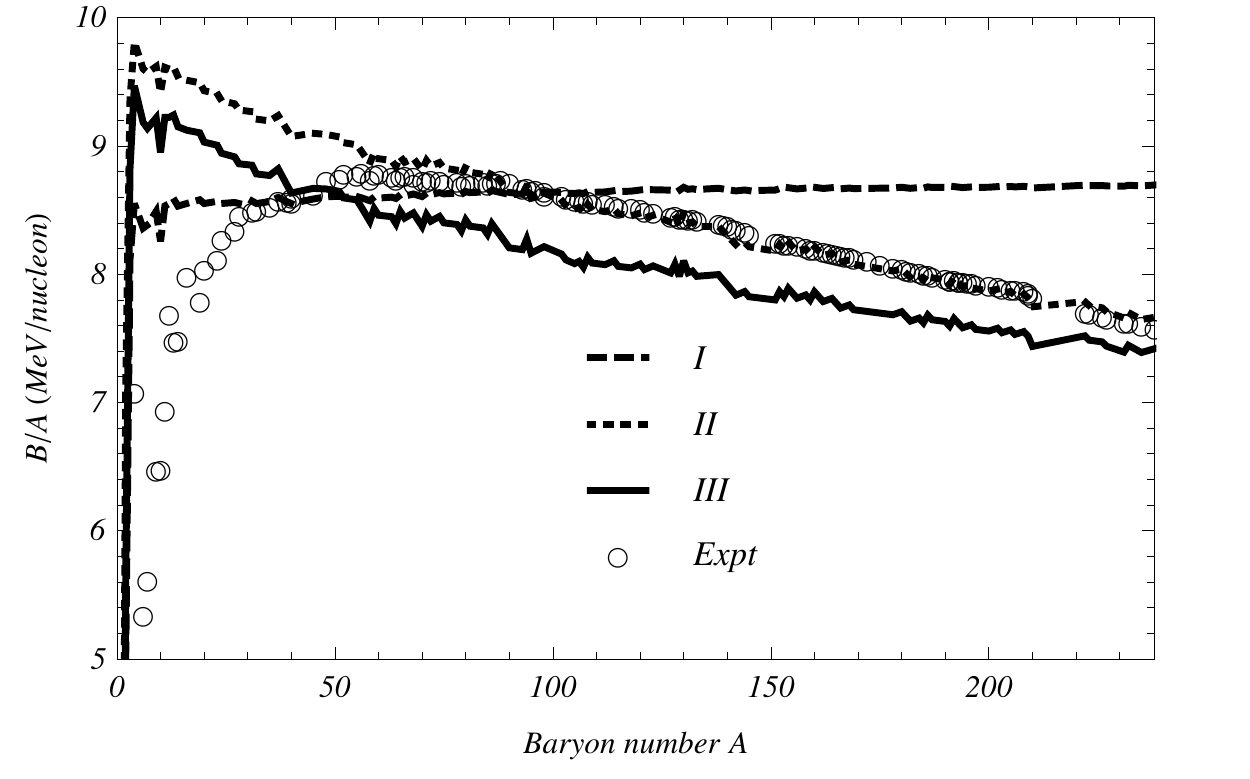}\caption{Binding
energy per nucleon $B/A$ as a function of the baryon number $A$: The
experimental data (empty circles) are shown along with predicted values for
parametrization of Set~I with $\alpha=\beta=0$ (dashed line), for Set~II, the
best fit for nuclear masses (dotted line), and for Set~III, the best fit for
$B/A$ (solid line), respectively.}%
\label{FigBoverA}%
\end{figure}

The results for $B/A$ are presented in Fig. \ref{FigBoverA} (dashed line).
They are compared to the experimental values (empty circles). We show here
only a subset of the table of nuclei in \cite{Audi:2002rp} composed of the
most abundant 140 isotopes. The parameters of Set~I lead to a sharp rise of
the binding energy per nucleon at small $A$ followed by a slow linear increase
for larger nuclei. The accuracy is found to be roughly within $10\%$ which is
relatively good considering the facts that the model involves only two
parameters at this point and the calculation involves a mass difference
between the nucleus and its constituents.

Experimentally the charge radius of the nucleus is known to behave
approximately as $\left\langle r_{\text{em}}^{2}\right\rangle ^{\frac{1}{2}%
}=r_{0}A^{\frac{1}{3}}$ with $r_{0}=1.23$~fm. It is straightforward to
calculate the root mean square radius of the baryon density [see Eq
(\ref{r2baryon})] which leads to $\left\langle r^{2}\right\rangle ^{\frac
{1}{2}}=\left(  2.007\text{~fm}\right)  A^{\frac{1}{3}}$. On the other hand
the charge radius $\left\langle r_{\text{em}}^{2}\right\rangle ^{\frac{1}{2}}$
displays a\ more complex dependence on $A$ since it involves an additional
isovector contribution (\ref{rhocharge})
\begin{equation}
\left\langle r_{\text{em}}^{2}\right\rangle =\frac{\int d^{3}r\ r^{2}%
\rho(\mathbf{r})}{\int d^{3}r\rho(\mathbf{r})}=\frac{A}{2Z}\left\langle
r^{2}\right\rangle +\frac{i_{3}}{Z}\left\langle r_{V}^{2}\right\rangle
\label{r2Z}%
\end{equation}
where $\rho(\mathbf{r})$ is given in expression (\ref{Ylm}) and $\left\langle
r_{V}^{2}\right\rangle $ is given by%
\[
\left\langle r_{V}^{2}\right\rangle =\frac{U_{11}^{(2)}}{a^{2}U_{11}}.
\]
where for the sake of conciseness we wrote $\left\langle r_{V}^{2}%
\right\rangle $ in terms of $U_{11}^{(2)}=U_{11}\left(  I_{lmn}^{2}\rightarrow
I_{lmn}^{4}\right)  .$ i.e. the integrals along the radial component in
$U_{11}^{(2)}$ contains an extra factor of $r^{2}$. Our computation verifies
that the charge radius obeys roughly the proportionality relation $\sim
r_{0}A^{\frac{1}{3}}$ but overestimates the experimental value of $r_{0}$ by
about $80\%$ with parameter Set~I.%

Let us now release the constraint $\alpha=\beta=0,$ and allow for small
perturbations from the nonlinear $\sigma$ and Skyrme term. In order to
estimate the magnitude of the parameters $\alpha$ and $\beta$ in a real
physical case, we perform two fits: the four parameters $\mu$, $\alpha,$
$\beta$ and $\lambda$ in Set~II optimizes the masses of the nuclei while
Set~III reaches the best agreement with respect to the binding energy per
nucleon, $B/A$. Both fits are performed with data from the same subset of the
most abundant 140 isotopes as before. The best fits on both cases would lead
to small negative values for $\beta$ similar to that of Refs.
\cite{Bonenfant:2010ab,Bonenfant:2012kt}. However, since the classical
(static) energy of the model is unbounded below if $\alpha,\beta<0$\ we impose
the constraint $\alpha,\beta\geq0$ from hereon to avoid stability problems.
(Note that in principle $\beta$ could take small negative values as long as
the Skyrme term is overcome by the repulsive Coulomb energy in which case the
physical nuclei would be stable but not the classical soliton.)

A summary of the results is presented in Table I while Fig. \ref{FigBoverA}
displays the general behavior of $B/A$ as a function of the baryon number for
Sets I, II, III, and experimental values. Note that the proton and neutron
mass differ slightly over Set I, II and III so for the sake of comparison we
use their experimental values in calculating $B/A.$

\[%
\begin{tabular}
[c]{|c|c|c|c|c|}\hline\hline
\multicolumn{5}{|c|}{Table I: Sets of parameters}\\\hline\hline
\  & $\quad$Set~I$\quad$ & $\quad$Set~II$\quad$ & $\quad$Set~III$\quad$ &
Expt.\\\hline\hline
$\mu$ $(10^{4}$ MeV$^{2})$ & $1.23223$ & $1.02259$ & $1.33515$ & ---\\
$\alpha$ $(10^{-3}$ MeV$^{2})$ & $0$ & $1.48244$ & $0.508933$ & ---\\
$\beta$ $(10^{-8}$ MeV$^{0})$ & $0$ & $1.20427$ & $1.31582$ & ---\\
$\lambda$ $(10^{-3}$ MeV$^{-1})$ & $4.74078$ & $5.70373$ & $4.36994$ & ---\\
$F_{\pi}$ $($ MeV$)$ & $0$ & $0.15401$ & $0.0902381$ & $186$\\
$m_{\pi}$ $($MeV$)$ & $0$ & $0$ & $0$ & $138$\\
$e^{2}$ ($10^{6}$) & --- & $2.59492$ & $2.37494$ & ---\\
$r_{0}$ (fm) & $2.00667$ & $2.27113$ & $1.90139$ & $1.23$\\\hline
\end{tabular}
\ \ \ \ \ \ \ \
\]

We find that the two new sets of parameters are very close to Set~I. In order
to make a relevant comparison, we look at the relative importance of the four
terms in (\ref{model0to6}) and how they scale with respect to the parameters
of the model, namely%
\[%
\begin{tabular}
[c]{rccccccc}
& $\mu\lambda$ & $:$ & $\alpha\left(  \lambda/\mu\right)  ^{1/3}$ & $:$ &
$\beta\left(  \mu/\lambda\right)  ^{1/3}$ & $:$ & $\mu\lambda$\\
$\text{Set~I\qquad}$ & $58.42$ & $:$ & $0$ & $:$ & $0$ & $:$ & $58.42$\\
$\text{Set~II\qquad}$ & $58.33$ & $:$ & $1.226\times10^{-5}$ & $:$ &
$1.463\times10^{-6}$ & $:$ & $58.33$\\
$\text{Set~III\qquad}$ & $58.35$ & $:$ & $3.507\times10^{-6}$ & $:$ &
$1.909\times10^{-6}$ & $:$ & $58.35$%
\end{tabular}
\ \ \
\]
for $\mathcal{L}_{0},\mathcal{L}_{2},\mathcal{L}_{4},$ and, $\mathcal{L}_{6},$
respectively. So the nonlinear $\sigma$ and Skyrme terms are found to be very
small compared to that of $\mathcal{L}_{0}\mathcal{\ }$and $\mathcal{L}_{6},$
i.e. by at least five orders of magnitude. This provides support to the
assumption that (\ref{FBeM}) is a good approximation to the exact solution.

The energy scale $\mu\lambda$ remain approximately the same for all the sets
while the values of $\mu$ and $\lambda$ shows noticeable differences. In
particular for the fit involving $B/A$ turns out to be somewhat sensitive to
these variations mostly because it involves a mass difference. We also note
some variation in the baryonic charge radius $r_{0}=1.3982\ \left(
\lambda/\mu\right)  ^{1/3};$ all sets overestimates the experimental value by
roughly 80\%. Since setting the parameters mainly involves fixing the relevant
energy scale $\mu\lambda$, perhaps the process may not be as sensitive to
setting a proper length scale for the nucleus so the predicted value of
$r_{0}$ should probably be taken as an estimate rather than a firm prediction.

Matching the parameters of the model with that of the original Skyrme Model,
we identify $F_{\pi}=4\sqrt{\alpha},\ e^{2}=1/32\beta$ whereas $m_{\pi}=0$ due
to the form of the potential$.$ The quantities $F_{\pi}$ and $e^{2}$ take
values which are orders of magnitude away for those obtained for the Skyrme
Model (see Table I) but this is not surprising since we have assumed from the
start that $\alpha$ and $\beta$ are relatively small. Unfortunately, one of
the successes of the original Skyrme Model is that it established a link with
soft-pion physics by providing realistic values for $F_{\pi}$, $m_{\pi}$ and
baryon masses. Such a link here is more obscure. The departure could come from
the fact that the parameters of the model are merely bare parameters and they
could differ significantly from their renormalized physical values. In other
words, we may have to consider two quite different sets of parameters: a first
one, relevant to the perturbative regime for pion physics where $F_{\pi}$ and
$m_{\pi}$ are closer to their experimental value and, a second set which
applies to the nonperturbative regime in the case of solitons. In our model,
this remains an open question.%

The model clearly improves the prediction of the nuclear masses and binding
energies in the regime where $\alpha$ and $\beta$ are small. Let us look more
closely at the results presented in Fig. \ref{FigBoverA}. The experimental
data (empty circles) are shown along with predicted values for parametrization
Set~I , Set~II and Set~III (dashed, dotted and solid lines, respectively).
Setting $\alpha=\beta=0$ (Set~I) leads to sharp increase $B/A$ at low baryon
number followed by a regular but slow growth in the heavy nuclei sector. This
suggests that heavier nuclei should be more stable, in contradiction to
observation. However the agreement remains within $\sim10\%$ in regards to the
prediction of the nuclear masses. This is significantly better than what is
obtained with the original Skyrme Model which overestimates $B/A$ by an order
of magnitude. Since $B/A$ depends on the difference between the mass of a
nucleus and that of its constituents, it is sensitive to small variation of
the nuclear masses so the results for $B/A$ may be considered as rather good.
The second fit (Set~II) is optimized for nuclear masses. The behavior at small
$A$ is similar to that of Set~I (as well as in Set~III) while it reproduces
almost exactly the remaining experimental values ($A\gtrsim40$). Finally, the
optimization of $B/A$ (Set~III) provide a somewhat better representation for
light nuclei at the expense of some of the accuracy found in Set~II for
$A\gtrsim40$. Overall, the binding energy is rather sensitive to the choice of
parameters. This is partly because the otherwise dominant contributions of
$E_{0}$ and $E_{6}$ to the total mass of the nucleus simply cancel out in
$B/A$.

The difference of behavior between light and heavy nuclei shown by the model
may be partly attributed to the (iso)rotational contribution to the mass. The
spin of the most abundant isotopes remains small while isospin can have
relatively large values due to the growing disequilibrium between the number
of proton and the number of neutron in heavy nuclei. On the other hand, the
moments of inertia increase with $A,$ so the total effect leads to a
(iso)rotational energy $E_{\text{r}}<$ 1 MeV with $A>10$ for all sets of
parameters considered and its contribution to $B/A$ decreases rapidly as $A$
increases. On the contrary, for $A<10,$ the rotational energy is responsible
for a larger part of the binding energy which means that $B/A$ \ should be
sensitive to the way the rotational energy is computed. So clearly, the
variations in shape of the baryon density has some bearing on the predictions
for the small $A$ sector not only the values of the parameters.

To summarize, the main purpose this work is to propose a model in a regime
where the nuclei are described by near-BPS solitons with approximately
constant baryon density configuration. This is acheived with a 4-terms
generalization of the Skyrme Model in the regime where the nonlinear $\sigma$
and Skyrme terms are considered small. The choice of an appropriate potential
$V$ allows to build constant baryon density near-BPS solitons, i.e. a more
realistic description of nuclei as opposed to the more complex configurations
found in most extensions of the Skyrme Model (e.g. $A=2$ toroidal , $A=3$
tetrahedral, $A=4$ cubic,...). Fitting the model parameters, we find a
remarkable agreement for the binding energy per nucleon $B/A$ with respect to
experimental data. On the other hand, there remain some caveats. First, the
Skyrme Model provides a simultaneous description for perturbative pion
interactions and nonperturbative baryon physics with realistic values for
$F_{\pi}$ and $m_{\pi}$ and baryon masses. The connection between the two
sectors here seems to be much more intricate. Secondly, there may be place for
improvement by proposing more appropriate solutions that would describe
equally well the light and heavy nuclei. Finally, the model seems unable to
reproduce a constant skin thickness in the baryon or charge density and the
experimental size of the nucleus correctly. On the other hand, the concept of
BPS-type Skyrmions also arises when one adds a large number of vector mesons
to the Skyrme Model as suggested by recent results based on holographic QCD
from Sutcliffe \cite{Sutcliffe:2010et}. Unfortunately, the emerging large $A$
Skyrmions configurations are rather complicated or simply unknown so that it
has yet been impossible to perform an analysis of the nuclear properties
comparable to that presented in this work. More recently Adam, Naya,
Sanchez-Guillen and Wereszczynski \cite{Adam:2013tda, *Adam:2013wya}
considered the special case of the pure BPS-model ($\alpha=\beta=0$) using the
potential $V_{\text{ASW}}$. Although their treatment differ slightly they find
a similar agreement for the binding energy per nucleon. Yet, all approaches
clearly suggest that nuclei could be treated as near-BPS Skyrmions.

This work was supported by the National Science and Engineering Research
Council of Canada.%

\bibliography{pub20131v6}%

\end{document}